\documentclass[onecolumn,showpacs,preprintnumbers,amsmath,amssymb]{revtex4}

\usepackage{graphicx}
\usepackage{dcolumn}
\usepackage{bm}

\begin{document}


\title{Superradiant scattering of electromagnetic waves emitted from
disk around Kerr black holes}

\author{Taichi Kobayashi}
 \email{kobayashi@gravity.phys.nagoya-u.ac.jp}
\author{Akira Tomimatsu}%
 \email{atomi@gravity.phys.nagoya-u.ac.jp}
\affiliation{
Department of Physics, Nagoya University, Chikusa-Ku, Nagoya, 464-8602, Japan
}%


\date{\today}

\begin{abstract}
 We study electromagnetic perturbations around a Kerr black hole surrounded by a thin disk on the equatorial plane. Our main purpose is to reveal the black hole superradiance of electromagnetic waves emitted from the disk surface. The outgoing Kerr-Schild field is used to describe the disk emission, and the superradiant scattering is represented by a vacuum wave field which is added to satisfy the ingoing condition on the horizon. The formula to calculate the energy flux on the disk surface is presented, and the energy transport in the disk-black hole system is investigated. Within the low-frequency approximation we find that the energy extracted from the rotating black hole is mainly transported back to the disk, and the energy spectrum of electromagnetic waves observed at infinity is also discussed.
\end{abstract}

\maketitle

\section{Introduction}

It is widely believed that there exists a rotating black hole surrounded by a disk in the central region of highly energetic astrophysical objects, such as active galactic nuclei, x-ray binary system and 
gamma-ray burst sources. In the disk-black hole system, the propagation of electromagnetic waves emitted from the disk surface will be strongly affected by the gravity of the central black hole. For example, the black hole shadow is expected to be an important phenomenon from which the black hole spin can be observationally estimated \cite{BHS1,BHS2,BHS3}.
The basic analysis of the shadow profile is relied on the ray-tracing method, and some features depending on various disk models have been investigated in \cite{RT1,RT2}. 
The further developments have been attempted in recent papers which take
into account the effects of plasma and wave scattering \cite{PERT,MCRT}.
Another key aspect of wave propagation in a rotating black hole
spacetime will be the effect of superradiance.
As was shown in \cite{ST,TP}, the amplitude of incident waves
propagating to the horizon can be amplified through the scattering
process to extract the rotational energy of a black hole.
Though this efficiency has been calculated for waves incident from
infinitely distant regions, in the disk-black hole system 
it is important to consider the electromagnetic waves 
which should occur on the disk surface.
In this paper we would like to assume a thin disk and deal with the
problem of propagation of waves incident from the equatorial plane in
the context of the superradiant scattering rather than from the
viewpoint of the black hole shadow.

The electromagnetic waves emitted from the disk surface should be
directly carried away to infinity, and partly absorbed by the black hole.
Our main interest is focused on the electromagnetic energy transport in
the disk-black hole system.
We will clarify how the wave absorption across the horizon generates
superradiant energy outflows from the black hole to disk and infinity.
This energy transport to the disk may be interpreted as a feedback mechanism
which plays a role of disk reheating, while the contribution to the
energy flux at infinity may be useful for an observational check of the
superradiance effect. (Such energy outflow may also contributes to a
change of the black hole shadow. However, to pursue this possibility is
beyond the scope of this paper.)

Because the ray-tracing method is not applicable to the analysis of the
superradiant scattering of electromagnetic waves, our approach to the
problem starts from solving the vacuum Maxwell equations in Kerr
geometry.
The assumed boundary condition is the existence of a thin disk (i.e., a
surface current) on the equatorial plane.
Namely, it is required that some components of electromagnetic fields
become discontinuous at the equatorial plane, and outgoing energy
fluxes are emitted from both the upper and lower sides of the disk surface.

The Kerr-Schild (K-S) formalism for solving the Einstein-Maxwell
equations is a useful method to overcome the mathematical difficulty due
to the disk boundary contribution \cite{DKS,TK1}.
In fact, if the electromagnetic fields are treated as perturbations in
Kerr geometry, all the field components are simply derived by two
arbitrary complex functions \cite{B2}, which can be appropriately chosen
according to the disk boundary condition.
Hence, in Sec.~\ref{KSNP}, we introduce the outgoing K-S field
$F_{\mu\nu}^{\rm KS}$ as a model of the disk emission and discuss a
singular behavior of the two complex functions at the equatorial plane.
Unfortunately, the outgoing K-S field fails to satisfy the condition
that no outgoing waves are present on the horizon.
Hence, the physical field satisfying the horizon boundary condition should be modified to the form
$F_{\mu\nu}=F^{\rm KS}_{\mu\nu}+F^{\rm SW}_{\mu\nu}$, 
where the additional vacuum field $F^{\rm SW}_{\mu\nu}$ be interpreted as the non-Kerr part
due to the superradiant  scattering of waves emitted from the disk and can be continuous even at 
the equatorial plane.
In order to facilitate the analysis of the scattering problem, 
we consider the Newman-Penrose quantities \cite{TP,NP} corresponding to
the electromagnetic field 
$F_{\mu\nu}=F_{\mu\nu}^{\rm KS}+F_{\mu\nu}^{\rm SW}$, and in
Sec.~\ref{Tf} we express them as the infinite sums of a complete set of
modes.
Based on the mode decomposition, in Sec.~\ref{Ef}, we derive the
formulae to calculate the energy fluxes at the boundary surfaces
including the horizon, the equatorial disk and infinity.
In Sec.~\ref{low}, the superradiant energy transport from the black hole
to the disk and infinity is explicitly estimated within the
low-frequency limit of the wave fields.
Hereafter we use units such that $c=G=1$.

\section{Kerr-Schild field and Newman-Penrose quantities\label{KSNP}}

Let us consider the electromagnetic waves emitted from disk surface
around Kerr black hole, using the framework of the Kerr-Schild formalism
(see details in \cite{DKS}).
Though this formalism is introduced to solve the full Einstein-Maxwell
equations, it may be applied to obtain electromagnetic perturbations on Kerr
background.
The metrical ansatz is 
\begin{equation}
 g^{\mu\nu}
  =
  \eta^{\mu\nu}-2He^{3\mu}e^{3\nu},
  \label{eq:KSform}
\end{equation}
where $\eta^{\mu\nu}$ is the metric of an auxiliary Minkowski spacetime, 
$H$ is scalar function,
and
$e^{3\mu}$ is a null vector field, which is tangent to a geodesic
and shear-free principal null congruence.
It is convenient to calculate the Einstein-Maxwell equations 
using tetrad components.
All other null tetrad
vectors are defined by the condition
\begin{equation}
 g_{ab}=e_{a}^{\ \mu}e_{b\mu}=
  \left(
   \begin{array}{cccc}
    0 & 1 & 0 & 0 \\
    1 & 0 & 0 & 0 \\
    0 & 0 & 0 & 1 \\
    0 & 0 & 1 & 0
   \end{array}
  \right)
  =g^{ab},
\end{equation}
where latin and greek suffixes mean tetrad  and tensor suffixes, respectively. 
A tensor $T_{\mu\ ...}^{\ \nu ...}$
is related to its tetrad components $T_{a\ ...}^{\ b ...}$
by either of the two equivalent relations
\begin{equation}
 T_{a\ ...}^{\ b...}=e_{a}^{\mu}e^{b}_{\nu}\ ...\ T_{\mu\ ...}^{\ \nu...},
  \
  T_{\mu\ ...}^{\ \nu ...}=e^{a}_{\mu}e_{b}^{\nu}\ ...\ T_{a\ ...}^{\ b...}.
\end{equation}

The essential point of the Kerr-Schild formalism is 
to use the complex form of electromagnetic field tensors
given by
\begin{equation}
 {\mathcal F}_{\mu\nu}
  \equiv F_{\mu\nu}+\frac{1}{2}i\epsilon_{\mu\nu\rho\sigma}F^{\rho\sigma},
\end{equation}
where $\epsilon_{\mu\nu\rho\sigma}$ is completely skew-symmetric, and
equal to $\epsilon_{1234}=(-g)^{1/2}$.
The corresponding null tetrad components are
\begin{equation}
 {\mathcal F}_{ab}=F_{ab}+\frac{1}{2}i\epsilon_{abcd}F^{cd},
  \label{eq:emtet}
\end{equation}
where $\epsilon_{abcd}$ is completely skew-symmetric, and
$\epsilon_{1234}=i$.
By virtue of the definition (\ref{eq:emtet}) and the Einstein equations,
the tetrad components ${\cal F}_{32}$, ${\cal F}_{41}$, and 
${\cal F}_{42}$ are found to be zero.
The electromagnetic fields are completely determined by only two complex
components ${\cal F}_{12}$, and ${\cal F}_{31}$. 
It is interesting to note that the Kerr-Schild form remains valid, even
if a back reaction on the gravitational field by the electromagnetic
field is considered.

A part of the Maxwell equations 
allows to write the tetrad components as
\begin{eqnarray}
 {\mathcal F}_{12}={\mathcal F}_{34}&=&AZ^{2},
  \label{eq:F1tet}\\
 {\mathcal F}_{31}&=&\gamma Z -(AZ)_{,1},
  \label{eq:F2tet}
\end{eqnarray}
where $Z$ is the complex expansion of 
the null vector $e^3$, and 
commas denote the directional derivatives along chosen
null tetrad vectors.
The functions $A$ and $\gamma$ should be
determined by solving the other Maxwell equations.

In this paper we treat the electromagnetic fields as perturbations on
Kerr background, and use the outgoing Kerr-Schild coordinate system to
describe electromagnetic waves emitted from disk to infinity.
The outgoing Kerr-Schild form of the Kerr metric is given by
\begin{eqnarray}
 {\mathrm d}s^2
  &=&
  -{\mathrm d}\tilde{t}^2
  +{\mathrm d}r^2
  +\Sigma {\mathrm d}\theta^2 
  +(r^2+a^2)\sin^2\theta {\mathrm d}\tilde{\varphi}^2
  \nonumber\\
 & & +2a\sin^2\theta {\mathrm d}r {\mathrm d}\tilde{\varphi}\nonumber\\
 & & +\frac{2Mr}{\Sigma}({\mathrm d}\tilde{t}-{\mathrm d}r-a
  \sin^2\theta{\mathrm d}\tilde{\varphi})^2,
\end{eqnarray}
where $\Sigma\equiv r^2+a^2\cos^2\theta$, and $M$ and $a$ denote the
mass and the angular momentum per unit mass of the black hole, respectively.
The function $H$ in Eq.~(\ref{eq:KSform}) is given by $2Mr/P^2\Sigma$,
where $P=1/\sqrt{2}\cos^2(\theta/2)$,
and the null tetrad vectors are given by
\begin{subequations}
\label{Eq.KStetrad}
\begin{eqnarray}
 e^{1}
  &=&
  2^{-\frac{1}{2}}e^{i\tilde{\varphi}}
  \left[
   -\tan(\theta/2),\
   \tan(\theta/2),\
   (r-ia\cos\theta),\
   (a+ir)\sin\theta
  \right],\\
 e^{2}
  &=&
  2^{-\frac{1}{2}}e^{-i\tilde{\varphi}}
  \left[
   -\tan(\theta/2),\
   \tan(\theta/2),\
   (r+ia\cos\theta),\
   (a-ir)\sin\theta
  \right],\\
 e^{3}
  &=&
  P
  \left[
   -1,\
   1,\
   0,\
   a\sin^2\theta
  \right],\\
 e^{4}
  &=&
  2^{-\frac{1}{2}}
  \left[
  1,\
  \cos\theta,\
  -r\sin\theta,\
  0
  \right]
  +
  He^{3},
\end{eqnarray}
\end{subequations}
with $e^{3}$ an outgoing null geodesic.
Then, the functions $A$ and $\gamma$ for electromagnetic perturbations
on Kerr background.
can be written as
\begin{eqnarray}
 A&=&\frac{\psi(Y,\tau)}{P^2}, \\
 \gamma&=&\frac{2^{1/2}\psi_{,\tau}}{P^2Y}+\frac{\phi(Y,\tau)}{P},
\end{eqnarray}
where 
$\tau=\tilde{t}-r+ia\cos\theta$ 
because $e^{3}$ is chosen as an outgoing vector field, 
$Y=e^{i\tilde{\varphi}}\tan(\theta/2)$ (see \cite{B2}), 
and comma means differentiation with respect to a given variable.
Further, we obtain $Z/P=1/(r-ia\cos\theta)$ for the complex expansion
$Z$ in Eqs.~(\ref{eq:F1tet}) and (\ref{eq:F2tet}). 
It should be noted that the tetrad components ${\mathcal F}_{ab}$ can written by 
the two arbitrary complex functions $\psi(Y,\tau)$ and $\phi(Y.\tau)$
as follows,
\begin{eqnarray}
 {\mathcal F}_{12}
  &=&
  \frac{\psi}{(r-ia\cos\theta)^2},
  \label{eq:f12}\\
 {\mathcal F}_{31}
  &=&
  \frac{1}{r-ia\cos\theta}
  \bigglb\{
   e^{-i\tilde{\varphi}}
   \left[
    \frac{2\cos^2(\theta/2)}{\tan(\theta/2)}
    \frac{r-ia}{r-ia\cos\theta}\psi_{,\tau}
    +\sin\theta\frac{r+ia}{(r-ia\cos\theta)^2}\psi
   \right]
   \nonumber \\ 
 &&+\phi(Y,\tau)
   -\frac{\psi_{,Y}}{r-ia\cos\theta}
  \bigglb\}.\label{eq:f31}
\end{eqnarray}
Hereafter we call this solution the Kerr-Schild field.

To see clearly superradiant energy transport in the
disk-black hole system,
it is convenient to introduce the Boyer-Lindquist coordinates, 
which lead to the metric
\begin{eqnarray}
 {\mathrm d}s^2
  &=&-\left(1-\frac{2Mr}{\Sigma}\right){\mathrm d}t^2
  -\frac{4aMr}{\Sigma}
  \sin^2\theta {\mathrm d}t {\mathrm d}\varphi
  +\frac{\Sigma}{\Delta}{\mathrm d}r^2
  +\Sigma {\mathrm d}\theta^2
  +\frac{{\mathcal A}}{\Sigma}\sin^2\theta {\mathrm d}\varphi^2,
\end{eqnarray}
where
$\Delta=r^2+a^2-2Mr$, and 
${\mathcal A}=(r^2+a^2)^2-a^2\Delta\sin^2\theta$.
The Boyer-Lindquist coordinates $t$ and $\varphi$ are related to
the outgoing Kerr-Schild coordinates $\tilde{t}$ and $\tilde{\varphi}$ as
follows,
\begin{eqnarray}
 {\mathrm d}t=
  {\mathrm d}\tilde{t}+\frac{2Mr}{\Delta}{\mathrm d}r,
  \label{eq:k-b1}\
 {\mathrm d}\varphi=
  {\mathrm d}\tilde{\varphi}+\frac{a}{\Delta}{\mathrm d}r.
  \label{eq:k-b2}
\end{eqnarray}

If the electromagnetic perturbations $F_{\mu\nu}$ written in the
Boyer-Lindquist coordinate system are assumed to be functions of three
variables $\sigma t-\varphi$, $r$, $\theta$ only, we can simply expect
that the superradiant scatterings occurs under the condition
$\sigma<\Omega_{\rm H}$ for the angular velocity $\Omega_{\rm H}$ of the
black hole and a frequency parameter $\sigma$ \cite{TK1}.
The expectation motivates us to specify the complex function
$\psi(Y,\tau)$ and $\phi(Y,\tau)$ in Eqs.~(\ref{eq:f12}) and 
(\ref{eq:f31}) to the forms
\begin{equation}
 \psi(Y,\tau)=\psi(X),\ 
  \phi(Y,\tau)\equiv (-ia\sigma X\psi_{,X}+\Psi(X))/Y.
\end{equation}
where the complex variable $X$ is defined by 
\begin{equation}
 X\equiv e^{-i\sigma\tau}Y
  =e^{-i\sigma\tau+i\tilde{\varphi}}\tan(\theta/2),
\end{equation}
and the term $- ia\sigma X\psi_{,X}$ is included in $\phi$ to simplify the
expression of $F_{\mu\nu}$ which will be given later.
It is easy to see from Eqs.~(\ref{eq:f12}) and (\ref{eq:f31}) that by
virtue of the choice of $\psi$ and $\phi$ the field components 
${\cal F}_{12}$ and ${\cal F}_{31}$ depend on $t$ and $\varphi$ via the
variable $\sigma\tau-\tilde{\varphi}$ in $X$, and we obtain
\begin{equation}
 \sigma\tau-\tilde{\varphi}=\sigma t-\varphi+i\sigma a\cos\theta
  -\sigma r_{*}+\Omega_{\rm H}L(r)
\end{equation}
where the tortoise coordinate $r_{*}$ is defined as
\begin{equation}
 r_{*}\equiv
  \int\frac{r^2+a^2}{\Delta}{\rm d}r
  =r-r_{1}+\frac{2Mr_{1}}{r_{1}-r_{2}}
  \ln\left|\frac{r-r_{1}}{r_{1}-r_{2}}\right|
  +\frac{2Mr_{2}}{r_{2}-r_{1}}
  \ln\left|\frac{r-r_{2}}{r_{1}-r_{2}}\right|,
\end{equation}
using the outer and inner horizon radii $r_{1}$ and $r_{2}$,
respectively.
For the function $L(r)$ given by
\begin{equation}
 L(r)=\int\frac{2Mr_{1}}{\Delta} {\mathrm d}r
  =\frac{2Mr_{1}}{r_{1}-r_{2}}\ln\left|\frac{r-r_{1}}{r-r_{2}}\right|,
\end{equation}
we can check the asymptotic behaviors such that $L-r_{*}\to 0$ on the
outer horizon $r=r_{1}$, and $L\to 0$ at infinity $r\to \infty$.

It is a straight forward task to derive the field components in the
Boyer-Lindquist coordinate system from ${\cal F}_{12}$ and 
${\cal F}_{31}$ given by 
Eqs.~(\ref{eq:f12}) and (\ref{eq:f31}).
Using the specified form of $\psi$ and $\phi$, the K-S field components
denoted by $F_{\mu\nu}^{\rm KS}$ are given by 
\begin{subequations}
\label{Eq.emboylin}
\begin{eqnarray}
 F_{tr}^{\rm KS}&=&-{\rm Re}
  \left[
   \frac{\psi(X)}{(r-ia\cos\theta)^2}
   -i\frac{a}{\Delta\sin\theta}
   \left(\Theta+i\Xi\right)
  \right],\\
 F_{t\theta}^{\rm KS}
  &=&
  -{\rm Re}
  \left[
   ia\sin\theta\frac{\psi(X)}{(r-ia\cos\theta)^2}
   +\frac{2}{\sin\theta}
   \left(\Theta+i\Xi\right)
  \right],\\
 F_{t\varphi}^{\rm KS}
  &=&-{\rm Re}
  \left[
   \frac{i}{\sin\theta}
   \left(\Theta+i\Xi\right)
  \right],\\
 F_{\theta\varphi}^{\rm KS}
  &=&{\rm Re}
  \left[
   -i(r^2+a^2)\sin\theta\frac{\psi(X)}{(r-ia\cos\theta)^2}
   +2a
   \left(\Theta+i\Xi\right)
  \right],\\
 F_{r\varphi}^{\rm KS}
  &=&{\rm Re}
  \left[
   -a\sin^2\theta\frac{\psi(X)}{(r-ia\cos\theta)^2}
   +i\frac{r^2+a^2}{\Delta\sin\theta}
   \left(\Theta+i\Xi\right)
  \right],\\
 F_{r\theta}^{\rm KS}
  &=&
  \frac{2\Sigma}{\Delta\sin\theta}
  {\rm Re}
  \left[
   \frac{1}{\sin\theta}
   \left(\Theta+i\Xi\right)
  \right],
\end{eqnarray}
\end{subequations}
where
the function $\Theta$ and $\Xi$ are defined as
\begin{eqnarray}
\Theta&\equiv&\Psi(X)+X\psi_{,X}
\left(2a\sigma\cos^2(\theta/2)-1\right)/(r-ia\cos\theta),\\
\Xi&\equiv& X\psi_{,X}a\sigma
 \left[2r\cos^2(\theta/2)/(r-ia\cos\theta)-1\right].
\end{eqnarray}

Here we consider the boundary condition on the disk located at
the equatorial plane $\theta=\pi/2$.
It is well-known that any complex function which is not a constant
should have a singularity on the complex plane.
We will assume that the existence of a singularity in $\psi(X)$ and
$\Psi(X)$ on
the complex $X$-plane, 
is due to a surface current on the equatorial plane $\theta=\pi/2$. 
This means that the components $F_{t\theta}^{\rm KS}$,
$F_{\theta\varphi}^{\rm KS}$,
 and $F_{r\theta}^{\rm KS}$ (namely, the imaginary part of $\psi$ and the real part of
$\Theta$) become discontinuous at $\theta=\pi/2$.
Such a discontinuity will be generated if a branch point in $\psi$
exists at $X=e^{i\beta}$ where $\beta$ is a real constant.
For example, as was discussed in \cite{TK1}, the function 
$\psi(X)=\psi_{0}\left[(X^{-2}+X^{2})^{3/2}-X^{-3}-X^3\right]^2$ with a
real constant $\psi_{0}$ has four branch points at $X^2=\pm i$, and the
imaginary part of $\psi$ and the real part of $X\psi_{,X}$ become
discontinuous at $\theta =\pi/2$. 
In this case the ratio $\Psi(X)/X\psi_{,X}$ may be chosen to be a real
constant,
as will be done in (\ref{Eq.cond_b}).
Further it should be noted
that the absolute value $|X|=e^{a\sigma\cos\theta}\tan(\theta/2)$
becomes equal to unity at $\theta=\pi/2$.
The branch point $X=e^{i\beta}$ may also appear on some conical plane 
$\theta=\theta_{0}(\neq \pi/2)$ giving $|X|=1$ if $a\sigma>1$.
Hence, in the following, the allowed range of the frequency parameter 
$\sigma$ is
limited to the range $0<\sigma<1/a$, for which we obtain $|X|<1$ in the upper
region $0\le\theta<\pi/2$ and $|X|>1$ in the lower region
$\pi/2<\theta\le \pi$.

We must also consider the regularity condition for $F_{\mu\nu}^{\rm KS}$ at the
polar axis (i.e., at $\theta=0$, $\pi$).
Noting that $|X|\simeq \sin(\theta/2)$ in the limit $\theta\to 0$
and $|X|\simeq 1/\cos(\theta/2)$ in the limit $\theta\to \pi$,
we find the boundary condition for $\psi$ and $\Psi$ to be 
for $\theta\to 0$,
\begin{equation}
 \psi(X)\sim \Psi(X)\sim X^{2},
  \label{Eq.condp1}
\end{equation}
and for $\theta\to \pi$,
\begin{equation}
 \psi(X)\sim  \Psi(X)\sim 1/X^2.
  \label{Eq.condp2}
\end{equation}

Finally, let us discuss the boundary condition on the 
horizon and at infinity,
by introducing
 the electromagnetic Newman-Penrose quantities $\phi_{a}$ ($a=0,\ 1,\ 2$)
\cite{NP}.
Using the Kinnersley's tetrad well-behaved on the past
horizon 
such that 
\begin{subequations}
\label{Eq.tetrad}
\begin{eqnarray}
 l^{\mu}&=&[(r^2+a^2)/\Delta,\ 1,\ 0,\ a/\Delta],\\
 n^{\mu}&=&[r^2+a^2,\ -\Delta,\ 0,\ a]/2\Sigma,\\
 m^{\mu}&=&[ia\sin\theta,\ 0,\ 1,\ i/\sin\theta]/2^{1/2}(r+ia\cos\theta),
\end{eqnarray}
\end{subequations}
the Newman-Penrose quantities are defined as
\begin{subequations}
 \label{Eq.emnp}
\begin{eqnarray}
 \phi_{0}&=&F_{\mu\nu}l^{\mu}m^{\nu},\\
 \phi_{1}&=&\frac{1}{2}F_{\mu\nu}
  (l^{\mu}n^{\nu}+\bar{m}^{\mu}m^{\nu}),\\
 \phi_{2}&=&F_{\mu\nu}\bar{m}^{\mu}n^{\nu},
\end{eqnarray}
\end{subequations}
and satisfy the Maxwell equations written by
\begin{subequations}
\label{Eq.maxnp}
\begin{eqnarray}
 \frac{(r-ia\cos\theta)}{\sqrt{2}}
  \left(
   {\cal L}_{1}-\frac{ia\sin\theta}{(r-ia\cos\theta)}
  \right)\phi_{0}
  =
  {\cal D}_{0}\left[
  \phi_{1}(r-ia\cos\theta)^2\right]\\
 \frac{1}{\sqrt{2}(r-ia\cos\theta)^2}
   {\cal L}_{0}\left[\phi_{1}(r-ia\cos\theta)^2\right]
  =
   {\cal D}_{0}\left[\phi_{2}(r-ia\cos\theta)\right],\\
 -\frac{\Delta}{\sqrt{2}(r-ia\cos\theta)^2}
   {\cal D}^{\dagger}_{0}\left[\phi_{1}(r-ia\cos\theta)^2\right]
  =
   {\cal L}^{\dagger}_{1}\left[\phi_{2}(r-ia\cos\theta)\right],
 \\
  -\frac{(r-ia\cos\theta)\Delta}{\sqrt{2}}
  \left(
   {\cal D}^{\dagger}_{1}-\frac{1}{(r-ia\cos\theta)}
  \right)\phi_{0}
  =
  {\cal L}^{\dagger}_{0}\left[\phi_{1}(r-ia\cos\theta)^2\right],
\end{eqnarray}
\end{subequations}
where the differential operators 
${\cal D}_{n}$ and ${\cal L}_{n}$ are defined by
\begin{subequations}
\begin{eqnarray}
 {\cal D}_{n}&\equiv&l^{\mu}\partial_{\mu}+2n(r-M)/\Delta,\\
 {\cal D}^{\dagger}_{n}
  &\equiv&-2(\Sigma/\Delta)n^{\mu}\partial_{\mu}+2n(r-M)/\Delta,\\
 {\cal L}_{n}&\equiv&
  \sqrt{2}(r-ia\cos\theta)\bar{m}^{\mu}\partial_{\mu}+n\cot\theta,\\
 {\cal L}^{\dagger}_{n}&\equiv&
  \sqrt{2}(r+ia\cos\theta)m^{\mu}\partial_{\mu}+n\cot\theta.
\end{eqnarray}
\end{subequations}
From Eqs.~(\ref{Eq.emboylin}), (\ref{Eq.tetrad}) and (\ref{Eq.emnp})
we obtain the Newman-Penrose quantities for the Kerr-Schild field
$F_{\mu\nu}^{\rm KS}$
as follows,
\begin{subequations}
\begin{eqnarray}
 \phi_{0}^{\rm KS}&=&0,\\
 \phi_{1}^{\rm KS}&=&\frac{\psi(X)}{2(r-ia\cos\theta)^2},\\
 \phi_{2}^{\rm KS}&=&
  \frac{1}{\sqrt{2}(r-ia\cos\theta)\sin\theta}
  \left[
   \Psi(X)
   +\frac{X\psi_{,X}}{(r-ia\cos\theta)}
   \left(
    i\sigma r\cos\theta +a\sigma-1
   \right)
  \right].\label{Eq.KS2}
\end{eqnarray}
\label{Eq.KSNP}
\end{subequations}
Because the electromagnetic waves are assumed to be emitted from the
disk, no ingoing waves should exist at infinity.
Hence, we require the asymptotic behavior of the Newman-Penrose
quantities obeying Eqs.~(\ref{Eq.maxnp}) to be
\begin{eqnarray}
 \phi_{0}\sim 1/r^3,\ \phi_{2}\simeq (1/r)g(t-r_{*}),
  \label{Eq.bound2}
\end{eqnarray}
in the limit $r\to \infty$. On the other hand no outgoing waves should
not exist on the horizon, and we require
\begin{equation}
 \phi_{0}\simeq (1/\Delta)f(t+r_{*}),\ \phi_{2}\sim \Delta,
  \label{Eq.boundhorizon}
\end{equation}
in the limit $\Delta\to 0$.
It is easy to see that the Kerr-Schild field 
satisfies the boundary condition only at infinity.
The horizon boundary condition breaks down, because $\phi_{2}^{\rm KS}$
does not vanish at $\Delta=0$.
Therefore, to obtain the physical field $\phi_{a}$ which is well-behaved
on the horizon, some vacuum field denoted by $\phi_{a}^{\rm SW}$ is
added to the Kerr-Schild field $\phi_{a}^{\rm KS}$ as follows
\begin{eqnarray}
 \phi_{0}=\phi_{0}^{\rm SW},\
  \phi_{1}=\phi_{1}^{\rm KS}+\phi_{1}^{\rm SW},\
 \phi_{2}=\phi_{2}^{\rm KS}+\phi_{2}^{\rm SW},
\end{eqnarray}
where $\phi_{2}$ is required to vanish on the horizon.
Though the Kerr-Schild field describes the disk emission,
the additional field is expected to represent the effect of wave
scattering (or absorption) by the black hole.
In the next section we will describe the scheme to obtain the additional
field $\phi_{a}^{\rm SW}$, by imposing the conditions (\ref{Eq.bound2}) 
and (\ref{Eq.boundhorizon})
on $\phi_{a}$.

\section{Wave scattering\label{Tf}}

As the first step to analyze the scattered wave field $\phi_{a}^{\rm SW}$,
let us expand the functions $\psi(X)$ and $\Psi(X)$ in Eqs.~(\ref{Eq.KSNP}) as
\begin{eqnarray}
 \psi(X)&=&\sum_{m}a_{m}X^{m},\label{Eq.exp_ps}\\
 \Psi(X)&=&\sum_{m}b_{m}X^{m},\label{Eq.exp_P}
\end{eqnarray}
where from the condition (\ref{Eq.condp1}) 
$m$ runs from $2$ to $\infty$ for $0<|X|<1$ (corresponding to the
upper region $0<\theta<\pi/2$), 
while from the condition (\ref{Eq.condp2}) 
it runs from $-2$ to $-\infty$ for 
$1<|X|<\infty$ (corresponding to the lower region $\pi/2<\theta<\pi$).
Such an expansion will be possible, because $\psi(X)$ and $\Psi(X)$ are
assumed to be regular except at branch points on the equatorial plane $|X|=1$.
Note that the $m$-th terms $\psi_{m}$ and $\Psi_{m}$ are proportional to 
$\exp\left[im(\sigma t-\varphi)\right]$, which represents a mode with
the wave frequency
\begin{equation}
 \omega_{m}\equiv m\sigma,
\end{equation}
for $m>0$, while the wave frequency should be understood to be $-\omega_{m}$ for $m<0$.
By virtue of the expansion of $\psi$ and $\Psi$,
the Kerr-Schild field $\phi_{2}^{\rm KS}$ is rewritten in to the form
\begin{eqnarray}
 (r-ia\cos\theta)^2\phi_{2}^{\rm KS}
  &=&
    \sum_{m=-\infty}^{\infty}
     e^{-i\omega_{m} t+im\varphi}
     S_{m}^{\rm KS}(r,\theta)e^{i(\omega_{m} r_{*}-m\Omega_{\rm H}L)},
     \label{Eq.KSp}\\
 S_{m}^{\rm KS}(r,\theta)
  &=&
  H_{m}
  \frac{e^{a\omega_{m}\cos\theta}\tan^{m}(\theta/2)}{\sqrt{2}\sin\theta}
  \left[
   (r-ia\cos\theta)b_{m}
   +a_{m}
   \left(
    i\omega_{m} r \cos\theta+a\omega_{m}-m
   \right)
  \right].\nonumber\\
  \label{Eq.KSp2}
\end{eqnarray}
Because the modes for $m=\pm1,\ 0$ should not be included in
Eq.(\ref{Eq.KSp}),
we have $H_{\pm 1}=H_{0}=0$.
Further the expansion forms (\ref{Eq.exp_ps}) and (\ref{Eq.exp_P}) mean
that 
for $m\ge 2$ (or $m\le -2$)
$\psi$ and $\Psi$ must vanish in the lower (or upper) region.
Hence, the factor $H_{m}$ in (\ref{Eq.KSp2}) is given by the step
function such that $H_{m}=(1+m/|m|)/2$ in the range $0\le \theta<\pi/2$ 
and $H_{m}=(1-m/|m|)/2$ in the range $\pi/2<\theta\le\pi$.
From Eq.~(\ref{Eq.KSp}) we have the asymptotic behavior near the horizon
$r=r_{1}$ as follows,
\begin{eqnarray}
 (r-ia\cos\theta)^{2}
  \phi_{2}^{\rm KS}
  \simeq
    \sum_{m}
     e^{-i\omega_{m} t+im\varphi}S_{m}^{\rm KS}(r_{1},\theta)e^{ik_{m}r_{*}}
     \label{Eq.p2KSh},\ k_{m}\equiv \omega_{m}-m\Omega_{\rm H}
\end{eqnarray}
which should be canceled out by the scattered-wave field 
$\phi_{2}^{\rm SW}$ according to the horizon boundary condition.
The easier way to construct such a vacuum non-Kerr-Schild field will be
to use the expansion form written by the spin-weighted angular functions
$S^{(\pm 1)}_{lm}(\theta)$.
The application of this mode decomposition to $\phi_{0}^{\rm SW}$ and 
$\phi_{2}^{\rm SW}$ leads to the result
\begin{eqnarray}
 \phi_{0}^{\rm SW}
  &=&
  \sum_{m,l}
  e^{-i\omega_{m} t+im\varphi}S^{(1)}_{lm}(\theta)R^{(1)}_{lm}(r),\\
 (r-ia\cos\theta)^2 \phi_{2}^{\rm SW}
  &=&
  \sum_{m,l}
  e^{-i\omega_{m} t+im\varphi}S^{(-1)}_{lm}(\theta)R^{(-1)}_{lm}(r),
\end{eqnarray}
where $R^{(s)}_{lm}$ is the radial function and $l\ge |m| \ge 2$.
Note that the component $\phi_{1}^{\rm SW}$
can be derived by using the Maxwell equations (\ref{Eq.maxnp}).
Therefore we consider hereafter only 
the two components $\phi_{0}^{\rm SW}$ and $\phi_{2}^{\rm SW}$.

The asymptotic behaviors of $R^{\pm 1}_{lm}$ near the horizon and at
infinity are well-known.
For example, in the limit $r\to r_{1}$,
we give the radial function as follows
\begin{eqnarray}
 R^{(1)}_{lm}&\simeq&
   C^{\rm In}_{lm}\Delta^{-1}e^{-ik_{m}r_{*}}
   +C^{\rm Out}_{lm}e^{ik_{m}r_{*}},\\
 R^{(-1)}_{lm}&\simeq&
   D^{\rm In}_{lm}\Delta e^{-ik_{m}r_{*}}
   +D^{\rm Out}_{lm}e^{ik_{m}r_{*}},
  \label{Eq.p2swh}
\end{eqnarray}
where the outgoing parts with the amplitudes $C_{lm}^{\rm Out}$ and 
$D_{lm}^{\rm Out}$ are also included to satisfy the condition
$\phi_{2}^{SW}+\phi_{2}^{\rm KS}\to 0$ on the horizon.
On the other hand, in the limit $r\to \infty$ where no incoming waves
exist
we obtain
\begin{eqnarray}
 R^{(1)}_{lm}&\simeq&
  E^{\rm Out}_{lm}e^{-i\omega_{m} r_{*}}/r^3,\\
 R^{(-1)}_{lm}&\simeq&
  F^{\rm Out}_{lm}re^{-i\omega_{m} r_{*}}.
  \label{Eq.infph2}
\end{eqnarray}
The coefficient ratios 
$C^{\rm In}_{lm}/D^{\rm In}_{lm}$ and 
$E_{lm}^{\rm Out}/F_{lm}^{\rm Out}$ 
have been derived in \cite{TP} using the Teukolsky equations,
and the results are given by
\begin{eqnarray}
 C^{\rm In}_{lm}/D^{\rm In}_{lm}&=&
  -\frac{32ik_{m}M^2r_{1}^{\ 2}(-ik_{m}+2\epsilon)}{B},
  \label{Eq.relDC}
  \\
 E^{\rm Out}_{lm}/F^{\rm Out}_{lm}&=&
  -\frac{B}{2\omega_{m}^{\ 2}},
\end{eqnarray}
where 
$B=(E+a^2\omega_{m}^{\ 2}
-2a\omega_{m}m)^2+4ma\omega_{m}-4a^2\omega_{m}^{\ 2}$ 
and $\epsilon=(r_{1}-r_{2})/4Mr_{1}$, and
$E$ is the eigenvalue of the angular equation.
In the low-frequency limit $a\omega_{m}\to 0$,
we have $E\to l(l+1)$, which is the case analyzed in Sec.~\ref{low}.
Further, we obtain 
the ratio	 $C^{\rm Out}_{lm}/D^{\rm Out}_{lm}$
written as
\begin{equation}
 C^{\rm Out}_{lm}/D^{\rm Out}_{lm}
  =\frac{B}{8ik_{m}M^2 r_{1}^{\ 2}(ik_{m}+2\epsilon)}.
  \label{Eq.relcd}
\end{equation}

From Eqs.~(\ref{Eq.p2KSh}) and (\ref{Eq.p2swh}),
the asymptotic behavior of  $\phi_{2}$ near the horizon
is written as 
\begin{equation}
 (r-ia\cos\theta)\phi_{2}
  \simeq
  \sum_{m,l}
  e^{-i\omega_{m} t+im\varphi}
  \left[
   S^{(-1)}_{lm}(\theta)
   \left(
    D_{lm}^{\rm In}\Delta e^{-ik_{m}r_{*}}+D_{lm}^{\rm Out}e^{ik_{m}r_{*}} 
   \right)
   +S_{m}^{\rm KS}(r_{1},\theta)e^{ik_{m}r_{*}}
  \right]
\end{equation}
Hence, the horizon boundary condition for $\phi_{2}$ leads to the relation
\begin{equation}
 \sum_{l\ge|m|}D^{\rm Out}_{lm}S^{(-1)}_{lm}(\theta)
  =-S^{\rm KS}_{m}(r_{1},\theta),
  \label{Eq.defD}
\end{equation}
from which the coefficient $D^{\rm Out}_{lm}$ is determined by
\begin{equation}
 D^{\rm Out}_{lm}
  \equiv-\int_{0}^{\pi}S^{\rm KS}_{m}(r_{1},\theta)S^{(-1)}_{lm}(\theta)\sin\theta
  {\rm d}\theta.
  \label{eq.detd}
\end{equation}
for the given Kerr-Schild field.

The important problem to be solved in relation to the superradiant
scattering of disk emission is to estimate the ratios
$C_{lm}^{\rm In}/D_{lm}^{\rm Out}$ and 
$F_{lm}^{\rm Out}/D_{lm}^{\rm Out}$, based on the radial equation \cite{TP}
\begin{equation}
 \Delta\frac{{\rm d}^{2}R^{(-1)}}{{\rm d}r^2}
  +\left[
    \frac{K^2+2i(r-M)K}{\Delta}-4ir\omega_{m} 
    -\lambda
   \right]
  R^{(-1)}=0,
  \label{Eq.radialeq}
\end{equation}
where $K\equiv (r^2+a^2)\omega_{m}-am$,
$\lambda$ is separation constant written by 
$\lambda=E-2am\omega_{m}+a^2\omega_{m}^{\ 2}$.

Finally, we summarize our proposed approach which is the derivation of
the scattered wave $F_{\mu\nu}^{\rm SW}$. 
In our approach, the property of the disk emission is given by the Kerr-Schild field 
$\phi_{a}^{\rm KS}$, 
that is, the two any complex functions $\psi$ and $\Psi$ or the
expansion coefficients $a_{m}$ and $b_{m}$ in Eqs.~(\ref{Eq.exp_ps}) and (\ref{Eq.exp_P}).
However the horizon boundary condition (\ref{Eq.boundhorizon}) breaks down, because
$\phi_{2}^{\rm KS}$ dose not vanish at $\Delta=0$.
Therefore, to obtain the physical field $\phi_{a}$ which is well-behaved
on the horizon,
we introduced the vacuum field $\phi_{a}^{\rm SW}$ which represents the
effect of wave scattering (or absorption) by the black hole. 
Then, the coefficient $D_{lm}^{\rm Out}$ in the outgoing part
of the scattered wave should be determined by Eq.~(\ref{eq.detd}),
using the Kerr-Schild field $\phi_{a}^{\rm KS}$. 
To determine the scattered field $\phi_{a}^{\rm SW}$,
we must solve the radial equation (\ref{Eq.radialeq}), using the boundary value
given by Eq.~(\ref{eq.detd}) on the horizon.
If the radial equation (\ref{Eq.radialeq}) is solved, 
the ratios 
$D_{lm}^{In}/D_{lm}^{\rm Out}$ and $F_{lm}^{\rm Out}/D_{lm}^{\rm Out}$ 
will be obtained.
Therefore, the scattered wave $\phi_{a}^{\rm SW}$
is represented only by the coefficients $D_{lm}^{\rm Out}$ connected with Kerr-Schild field
on the horizon.

Before pursuing the analysis of Eq.~(\ref{Eq.radialeq}) in details, we must present the
formulae to calculate the energy fluxes from the disk,
on the horizon and at infinity, because our main purpose is to clarify
the energy transport via the superradiant scattering process in the
disk-black hole system.
This will be done in the next section.

\section{Energy flux\label{Ef}}

Using the Newman-Penrose quantities obtained in the previous section,
let us present the useful expressions of the energy
flux vector defined by 
\begin{equation}
 {\cal E}^{\mu}\equiv-T^{\mu}_{\ t},
  \label{Eq.defen}
\end{equation}
where
\begin{eqnarray}
 T_{\mu\nu}
  &=&
  \frac{1}{4\pi}
  \left[
   \phi_{0}\bar{\phi}_{0}n_{\mu}n_{\nu}
   +\phi_{2}\bar{\phi}_{2}l_{\mu}l_{\nu}
   +2\phi_{1}\bar{\phi}_{1}\left(l_{(i}n_{j)}+m_{(i}\bar{m}_{j)}\right)
   \right.\nonumber \\ &&
  \left.
   -4\bar{\phi}_{0}\phi_{1}n_{(i}m_{j)}
   -4\bar{\phi}_{1}\phi_{2}l_{(i}m_{j)}
   +2\phi_{2}\bar{\phi}_{0}m_{i}m_{j}
  \right]
  +{\rm C.C.}.
  \label{Eq.EneT}
\end{eqnarray}
Because the component $\phi_{1}$ given by $\phi_{0}$ and $\phi_{2}$ 
through the Maxwell equations (\ref{Eq.maxnp}), 
 the energy flux vector ${\cal E}^{\mu}$
can be written by $\phi_{0}$ and $\phi_{2}$ only.

Note that the energy flux vector for the wave fields considered here
is oscillatory with respect to the
time $t$ (as well as the azimuthal angle $\varphi$).
To estimate the efficiency of the energy transport,
we must consider the time-average quantities such that
\begin{equation}
 \langle A\rangle
  \equiv \frac{\sigma}{2\pi}
  \int_{0}^{2\pi/\sigma}
  A {\rm d}t,
  \label{Eq.t-aq}
\end{equation}
with the frequency parameter $\sigma$.
Because it is easy to see that
the time-dependence on the energy flux vectors ${\cal E}^{\mu}$ 
arises from the electromagnetic field components $\phi_{a}$,
we consider the time-averaged quantities $\langle\phi_{a}\bar{\phi}_{b}\rangle$ 
written as
\begin{equation}
  \langle \phi_{a}\bar{\phi}_{b}\rangle
  = \frac{\sigma}{2\pi}
  \int_{0}^{2\pi/\sigma}
  \phi_{a}\bar{\phi}_{b} {\rm d}t.
\end{equation}
Note that 
the electromagnetic field components $\phi_{a}$ are expanded as
\begin{equation}
 \phi_{a}=\sum_{m=-\infty}^{\infty}\phi_{am},
\end{equation}
with $|m|\ge 2$.
In particular, $\phi_{0m}$ and $\phi_{2m}$ are written as
\begin{eqnarray}
 \phi_{0m}&=&e^{-i\omega_{m} t+im\varphi}
  \sum_{l\ge|m|}S^{(1)}_{lm}(\theta)R^{(1)}_{lm}(r),
  \label{Eq.phi0m}\\
 (r-ia\cos\theta)^2\phi_{2m}&=&e^{-i\omega_{m} t+im\varphi}
  \left[
   \sum_{l\ge|m|}S^{(-1)}_{lm}(\theta)R^{(-1)}_{lm}(r)
   +S_{m}^{\rm KS}(r,\theta)e^{im\left(\sigma r_{*}-\Omega_{\rm H}L(r)\right)}
  \right].
  \label{Eq.phi2m}
\end{eqnarray}
From Eqs.~(\ref{Eq.phi0m}) and (\ref{Eq.phi2m}),
it is easy to see that 
the time-averaged quantities $\langle \phi_{a}\bar{\phi}_{b}\rangle$ are given by
the mode decompositionas follows,
\begin{equation}
 \langle \phi_{a}\bar{\phi}_{b}\rangle
  =\sum_{m=-\infty}^{\infty}\phi_{a m}\bar{\phi}_{b m},
\end{equation}
with $|m|\ge 2$,
which will lead to 
the time-averaged energy flux vectors written as 
\begin{equation}
 \langle{\cal E}^{\mu}\rangle
  =\sum_{m=-\infty}^{\infty}
  \langle{\cal E}^{\mu}_{m}\rangle,
  \label{Eq.expene}
\end{equation}
with $|m|\ge 2$.
Hereafter, we consider the mode-decomposed and time-averaged energy
flux vector $\langle{\cal E}^{\mu}_{m}\rangle$.

We consider the angular component $\langle{\cal E}^{\theta}\rangle$ of
the energy flux vector as the emission from the disk surface.
Nothing that $\sqrt{\Sigma}=r$ at $\theta=\pi/2$, the energy flux 
${\cal E}_{\rm D}$ per unit area can be evaluated as
\begin{equation}
 {\cal E}^{\rm D}_{m}(r)=
  r\langle{\cal E}^{+}_{m}\rangle
  +r\langle{\cal E}^{-}_{m}\rangle,
\end{equation}
where $\langle{\cal E}^{\pm}_{m}\rangle$
are equal to 
$\pm\langle{\cal E}^{\theta}_{m}\rangle$ in the limit 
$\theta\to \pi/2 \pm 0$ corresponding to the disk emission from the 
upper and lower side, respectively.
Then, $\langle{\cal E}^{\pm}_{m}\rangle$ are obtained as
\begin{eqnarray}
 \langle{\cal E}_{m}^{\pm}\rangle
  &=&
   -\frac{1}{8\pi r^2 (m-a\omega_{m})}\partial_{r}A_{m}^{\pm}
   -\frac{\omega_{m}}{4\pi (m-a\omega_{m})}iB_{m}^{\pm}
   ,\\
 A^{\pm}_{m}&=&
   \left\{
    \Delta
    (\bar{\phi}^{\pm}_{0m}\phi^{\pm}_{2m}
    +\phi^{\pm}_{0m}\bar{\phi}^{\pm}_{2m})
    +2r^2 \phi^{\pm}_{2m}\bar{\phi}^{\pm}_{2m}
    +\frac{\Delta^2}{2r^2}\phi^{\pm}_{0m}\bar{\phi}^{\pm}_{0m}
   \right\},\\
 B^{\pm}_{m}&=&
   (\bar{\phi}^{\pm}_{0m}\phi^{\pm}_{2m}-
   \phi^{\pm}_{0m}\bar{\phi}^{\pm}_{2m}).
\end{eqnarray}
To evaluate the total flux radiated from disk surface,
it is easy to obtain the total energy flux $E^{\rm D}_{m}$ as follows
\begin{eqnarray}
 E^{{\rm D}}_{m}&\equiv&
  2\pi\int_{r_1}^{\infty}
  {\cal E}^{\rm D}_{m}r{\rm d}r.
  \label{Eq.EneDT}
\end{eqnarray}
After a tedious calculation, we have the total energy flux $E^{\rm D}_{m}$ 
given by
\begin{eqnarray}
  E^{{\rm D}}_{m}
   &=&
   -\frac{1}{4(m-a\omega_{m})}
   \left[A^{+}_{m}-A^{-}_{m}\right]^{\infty}_{r_1}
   -\frac{\omega_{m}}{2(m-a\omega_{m})}\int_{r_1}^{\infty}
   i(B^{+}_{m}-B^{-}_{m})r^2 {\rm d} r.
\end{eqnarray}

Noting that for the Kerr-Schild field
$\phi_{0}^{\rm KS}=0$ and $\phi_{2}^{\rm KS}\to 0$ in the limit $\theta\to\pi/2-0$,
the corresponding total energy flux 
$[E^{{\rm D}}_{m}]^{\rm KS}$ is obtained as
\begin{eqnarray}
 [E^{{\rm D}}_{m}]^{\rm KS}&=&
  \frac{1}{2(m-a\omega_{m})}
  \left[
   (r^2\phi_{2m}^{+\rm KS}\bar{\phi}_{2m}^{+\rm KS})_{\rm H}
   -(r^2\phi_{2m}^{+\rm KS}\bar{\phi}_{2m}^{+\rm KS})_{\infty}
  \right].
\end{eqnarray}
To assume the continuity of $\phi_{a}^{\rm SW}$ at $\theta=\pi/2$ on the
horizon for $m\ge 2$,
we impose the same continuity condition $\phi_{a}^{\rm KS}$
which is interpreted as the condition that the disk 
does not extend to the horizon,
and choose
$(r^2\phi_{2m}^{+\rm KS}\bar{\phi}_{2m}^{+\rm KS})_{\rm H}$
to be zero.
Then, from Eq.~(\ref{Eq.p2KSh}) it is easy to see that
the coefficient $b_{m}$ is determined as follows,
\begin{equation}
 b_{m}=\frac{m-a\omega_{m}}{r_{1}}a_{m}.\label{Eq.cond_b}
\end{equation}
In particular, 
noting the continuity of the scattered field at $\theta=\pi/2$, namely,
$\phi_{a}^{+{\rm SW}}=\phi_{a}^{-{\rm SW}}$,
the total energy flux $E_{{\rm D}m}$ is obtained as follows,
\begin{eqnarray}
 E^{{\rm D}}_{m}
  &=&
  -\frac{(r^2\phi_{2m}^{+\rm KS}
  \bar{\phi}_{2m}^{+\rm KS})_{\infty}}{2(m-a\omega_{m})}
  \nonumber\\&&
  -\frac{(r^2 \phi_{2m}^{+{\rm KS}}\bar{\phi}_{2m}^{+{\rm SW}})_{\infty}}
  {2(m-a\omega_{m})}
  +\frac{\omega_{m}}{m-a\omega_{m}}\int_{r_1}^{\infty}
  {\rm Im}(\phi_{2m}^{+{\rm KS}}\phi_{0m}^{+{\rm SW}})r^2{\rm d}r.
\end{eqnarray}
For $m\le -2$,
we can use the same formula
only by exchanging the subscript $+$ for $-$ and vice versa.



Next 
let us calculate
the energy flux on the horizon,
where we obtain the radial component 
$\langle{\cal E}^{r}_{m}\rangle$ of the energy flux vector as
\begin{eqnarray}
  \langle{\cal E}^{r}_{m}\rangle_{\rm H}
   =
  -\frac{1}{8\pi k_{m}\Sigma_{\rm H}}
  \left[
   \frac{\omega_{m}}{2Mr_{1}}
   (\Delta^2\phi_{0m}\bar{\phi}_{0m})_{\rm H}
   -\frac{\Omega_{\rm H}}{2\sin\theta}
   \partial_{\theta}
   \left(
    \frac{(\Delta^2\phi_{0m}\bar{\phi}_{0m})_{\rm H}}{\Sigma_{\rm H}}\sin^2\theta
   \right)
  \right].
  \label{Eq.Erh}
\end{eqnarray}
Further,
we can evaluate the total flux $E_{{\rm H}}$ integrated over the whole
horizon
surface as follows,
\begin{eqnarray}
  E^{{\rm H}}_{m}
  &\equiv&2\pi\int_{0}^{\pi}\langle{\cal E}^{r}_{m}\rangle_{\rm H}
  \Sigma_{\rm H}\sin\theta{\rm d}\theta,
  \label{Eq.intener}
\end{eqnarray}
which reduces to the form
\begin{eqnarray}
 E^{{\rm H}}_{m}&=&
  -\frac{\omega_{m}}{8k_{m}Mr_{1}}\int_{0}^{\pi}
  \left(\Delta^2\phi_{0m}\bar{\phi}_{0m}\right)_{\rm H}\sin\theta 
  {\rm d}\theta.
  \label{Eq.htotal}
\end{eqnarray}
Note that the integration of  the second term in Eq.~(\ref{Eq.Erh}) 
is canceled out, because $\phi_{0m}$ is continuous even at $\theta=\pi/2$.
The result given by Eq.~(\ref{Eq.htotal}) shows that
the energy extraction from the black hole occurs for
incident waves with the frequency parameter $\sigma$ in the range 
$0<|\omega_{m}|<|m|\Omega_{\rm H}$ 
(i.e., $k_{m}<0$  for $m<0$ and 
 $k_{m}>0$ for $m>0$) which means the range
$0<\sigma<\Omega_{\rm H}$,
in accordance with the result of the usual
superradiant scattering \cite{TP}.
We can rewrite the net flux $E^{{\rm H}}_{m}$
into the form
\begin{equation}
  E^{{\rm H}}_{m}
  =
  -\frac{\omega_{m}}{8k_{m}Mr_{1}}
  \sum_{l}|C^{\rm In}_{lm}|^2.
  \label{Eq.Tform}
\end{equation}
in which the coefficient $C^{\rm In}_{lm}$ will be given by solving the
radial equation (\ref{Eq.radialeq}) under the low-frequency limit (see Sec.~\ref{low}).

Finally we turn our attention to the radial component of energy flux vector 
$\langle{\cal E}^{r}\rangle_{m}$
at infinity.
The energy flux vector is written by 
\begin{eqnarray}
 \langle{\cal E}^{r}_{m}\rangle_{\infty}
  &=&
  \frac{1}{2\pi r^2}(r^2\phi_{2m}\bar{\phi}_{2m})_{\infty}.
\end{eqnarray}
It is easy to see that no ingoing energy flux exists at infinity.
Further we calculate the total flux at infinity as follows,
\begin{equation}
 E^{\infty}_{m}
  \equiv
  2\pi\int_{0}^{\pi}\langle{\cal E}^{r}_{m}\rangle_{\infty} r^2\sin\theta
  {\rm d}\theta.
  \label{Eq.EneinfT}
\end{equation}
Then
the total flux at infinity is given by
\begin{eqnarray}
 E^{\infty}_{m}&=&
  \int_{0}^{\pi}(r^2 \phi_{2m}\bar{\phi}_{2m})\sin\theta
  {\rm d}\theta.
\end{eqnarray}
Considering the contributions form the 
Kerr-Schild field and the scattered wave,
the total flux at infinity is rewritten by
\begin{equation}
 E^{\infty}_{m}=
 \int_{0}^{\pi}
  r^2\left[
   |\phi_{2m}^{\rm KS}|^2
   +2{\rm Re}(\phi_{2m}^{\rm KS}\bar{\phi}_{2m}^{\rm SW})
   +|\phi_{2m}^{\rm SW}|^2
  \right]_{\infty}
  \sin\theta{\rm d}\theta,
  \label{Eq.einf}
\end{equation}
where from Eqs.~(\ref{Eq.KS2}), (\ref{Eq.exp_P}), and (\ref{Eq.cond_b}) 
we obtain
 the asymptotic form of Kerr-Schild field as
\begin{equation}
 \phi_{2m}^{\rm KS}
  \to
  -e^{-i\omega_{m} t+im\varphi+i\omega_{m} r_{*}}
  H_{m}\frac{a_{m}e^{a\omega_{m}\cos\theta}\tan^{m}(\theta/2)}
  {\sqrt{2}r r_1\sin\theta}
  \left[
   (m-a\omega_{m})-i\omega_{m} r_1\cos\theta
  \right],
  \label{Eq.ksinf}
\end{equation}
while $\phi_{2m}^{\rm SW}$ has the asymptotic
form given by  Eq.~(\ref{Eq.infph2}).

Here, we assume the Kerr-Schild field to be reflection-symmetric
with respect to the equatorial plane.
This symmetry is corresponding to the condition that the two expansion coefficients
are written by
\begin{equation}
 a_{m}=a_{-m},\ b_{m}=-b_{-m},
\end{equation}
which allows us to 
calculate the energy flux for $m<0$, by using the result for $m>0$.

In following section,
we see the energy transport to calculate the total energy in each region
using the low-frequency limit.
Then we should solve the radial equation (\ref{Eq.radialeq}) 
to determine the amplitude of the electromagnetic fields on the horizon and at infinity.

\section{Low-frequency limit\label{low}}

In the previous section,
we discussed the time-averaged energy flux vector on the boundary surface.
To evaluate explicitly the efficiency of the superradiant scattering,
we attempt to solve the radial equation (\ref{Eq.radialeq}) using the low-frequency limit
($a\omega_{m}\to 0$) according to the procedure in \cite{ST}.

For $a\omega_{m}\ll 1$,
the angular function $S^{\pm1}_{lm}(\theta)$ is known to be given by 
\begin{eqnarray}
 S^{(s)}_{lm}(\theta)
  &=&
  (-1)^{m}
  \sqrt{\frac{2l+1}{2}\frac{(l+m)!}{(l+s)!}\frac{(l-m)!}{(l-s)!}}
  \sin^{2l}(\theta/2)
  \nonumber\\ && \times 
  \sum_{r=0}^{l-s}
  \left[
   \left(
    \begin{array}{c} l-s\\ r\end{array}
   \right)
   \left(
    \begin{array}{c} l+s\\ r+s-m\end{array}
   \right)
   (-1)^{l-r-s}\cot^{2r+s-m}(\theta/2)
  \right],
\end{eqnarray}
where the eigenvalue $E$ is obtained by 
$E=(l-s)(l+s+1)$ (see \cite{SSphe}).
On the other hand, we obtain the radial function through the asymptotic
matching method.
To consider the dimenssionless radial equation from 
Eq.~(\ref{Eq.radialeq}),
we introduce the dimenssionless condition and parameter as
\begin{eqnarray}
 x&=&\frac{r-r_{1}}{r_{1}-r_{2}},\\
 Q&=&\frac{r_{1}^{\ 2}+r_{2}^{\ 2}}{r_{1}-r_{2}}(m\Omega_{\rm H}-\omega_{m}).
\end{eqnarray}
Then,
for $\omega_{m} M\ll 1$ and $x\gg {\rm max}(Q,l)$ (i.e., in the distant region ),
the radial equation can be rewritten by
\begin{equation}
 \frac{{\rm d}^2 R^{(-1)}_{lm}}{{\rm d}x^2}
  +\left[
    \omega_{m}^2 (r_{1}-r_{2})^{2}
    -
    \frac{2i\omega_{m}(r_{1}-r_{2})}{x}
    -\frac{l(l+1)}{x^2}
  \right]R^{(-1)}_{lm}=0.
\end{equation}
This equation is expressible in terms of the confluent hypergeometric
functions,
and we obtain the outgoing wave solution given by
\begin{equation}
 R^{(-1)}_{lm}
  =C \frac{2\kappa xe^{-\kappa x}}{(l-1)!}
  \int_{0}^{\infty}
  e^{-t}t^{l-1}
  \left(1+\frac{t}{2\kappa x}\right)^{l+1}{\rm d}t,
  \label{Eq.farsol}
\end{equation}
with $\kappa  =- i(r_{1}-r_{2})\omega_{m}$, and $C=F_{lm}^{\rm out}/(2i\omega_{m})$.
It is easy to see that
the asymptotic behavior of Eq.~(\ref{Eq.farsol}) is written by
written by
\begin{equation}
 R^{(-1)}_{lm}
  \simeq C 2\kappa xe^{-\kappa x},
\end{equation}
for $|\kappa x|\gg 1$ in accordance with Eq.~(\ref{Eq.infph2}),
and 
\begin{equation}
 R^{(-1)}_{lm}
  \simeq C\frac{(2l)!}{(l-1)!}(2\kappa)^{-l-1}x^{-l},
  \label{Eq.asyf}
\end{equation}
for $|\kappa x|\ll 1$.
In the region near the horizon ($x\ll l/\omega_{m} (r_{1}-r_{2})$)
the radial equation (\ref{Eq.radialeq}) can be rewritten as
\begin{equation}
 [x(x+1)]^{2}\frac{{\rm d}^2 R^{(-1)}_{lm}}{{\rm d}x^2}
  +
  \left[
   Q^2-iQ(1+2x)-l(l+1)x(x+1)
  \right]
  R^{(-1)}_{lm}=0,
  \label{Eq.ndradeq}
\end{equation}
where it is obtained by neglecting in
Eq.~(\ref{Eq.radialeq}) all the terms containing $\omega_{m}$ except the one
which enters into $Q$, and the solution with ingoing and outgoing
boundary condition at the horizon is given by
\begin{equation}
 R^{(-1)}_{lm}
  =R^{(-1)}_{lm}({\rm In}) +R^{(-1)}_{lm}({\rm Out}),
  \label{Eq.nearhr}
\end{equation}
where $R^{(-1)}_{lm}({\rm In})$ and $R^{(-1)}_{lm}({\rm Out})$ are written by
\begin{eqnarray}
 R^{(1)}_{lm}({\rm In})&=&
  -A\frac{(2l+1)!}{(l-1)!}\frac{\Gamma(-2iQ)}{\Gamma(2+2iQ)}
  \frac{\Gamma(1+2iQ)}{\Gamma(l+1-2iQ)}
  \left(\frac{x}{x+1}\right)^{iQ}x(x+1)\nonumber\\&&\times
  \frac{\Gamma(2+2iQ)}{\Gamma(1-l)(l+1)!}
  \sum_{n=0}^{l-2}\frac{\Gamma(n+1-l)\Gamma(n+l-1)}{\Gamma(n+2+2iQ)}
  \frac{(-x)^{n}}{n!}
  ,
  \label{Eq.RIn}\\
 R^{(-1)}_{lm}({\rm Out})&=&
  A\frac{(2l+1)!}{(l+1)!}
  \frac{\Gamma(1+2iQ)}{\Gamma(l+1+2iQ)}
  \left(\frac{x}{x+1}\right)^{iQ}\nonumber\\&&\times
  \frac{\Gamma(2iQ)}{\Gamma(-1-l)(l-1)!}
  \sum_{n=0}^{l}\frac{\Gamma(n-1-l)\Gamma(n+l)}{\Gamma(n+2iQ)}
  \frac{(-x)^{n}}{n!},
  \label{Eq.ROut}
\end{eqnarray}
respectively.
From Eqs.~(\ref{Eq.RIn}) and (\ref{Eq.ROut}),
the amplitudes $D_{lm}^{\rm In}$ and $D_{lm}^{\rm Out}$ 
are given by
\begin{eqnarray}
 D_{lm}^{\rm In}
  &=&
  -(r_{1}-r_{2})^2A\frac{(2l+1)!}{(l-1)!}\frac{\Gamma(-2iQ)}{\Gamma(2+2iQ)}
  \frac{\Gamma(1+2iQ)}{\Gamma(l+1-2iQ)}
  ,\\
 D_{lm}^{\rm Out}
  &=&
  A\frac{(2l+1)!}{(l+1)!}
  \frac{\Gamma(1+2iQ)}{\Gamma(l+1+2iQ)}.
\end{eqnarray}
We obtain the solution with the asymptotic behavior  as follows,
\begin{equation}
 R^{(-1)}_{lm}
  \simeq
  (-1)^{-l}Ax^{-l},
  \label{Eq.asyn}
\end{equation}
for $x\gg 1$.
If all the parameters satisfy the condition
${\rm max}(Q,l)\ll  l/\omega_{m} (r_{1}-r_{2})$,
we have the overlap region in which both the outer expression
(\ref{Eq.asyf})
and the inner expression (\ref{Eq.asyn}) hold.
In this region,
we can match the leading-terms of the solutions (\ref{Eq.asyf}) and
(\ref{Eq.asyn}), and this matching yields
\begin{equation}
 A=(-1)^{l}(2k)^{-l-1}\frac{(2l)!}{(l+s)!}C.
\end{equation}
Then we can obtain the ratios
${F_{lm}^{\rm Out}}/{D^{\rm Out}_{lm}}$ and 
${D_{lm}^{\rm In}}/{D^{\rm Out}_{lm}}$ as follows,
\begin{eqnarray}
 \frac{F_{lm}^{\rm Out}}{D^{\rm Out}_{lm}}
  &=&(-1)^{-l}(r_{1}-r_{2})^{-1}(2\kappa )^{l+1}
  \frac{(l-1)!(l+1)!}{(2l)!(2l+1)!}
  \frac{\Gamma(l+1+2iQ)}{\Gamma(1+2iQ)},
  \label{Eq.relhinf}\\
 \frac{D_{lm}^{\rm In}}{D^{\rm Out}_{lm}}
  &=&-\frac{1}{(r_{1}-r_{2})^{2}}
  \frac{(l+1)!}{(l-1)!}
  \frac{\Gamma(l+1+2iQ)\Gamma(-2iQ)}{\Gamma(l+1-2iQ)\Gamma(2+2iQ)}.
  \label{Eq.relDD}
\end{eqnarray}
Here the coefficients  
${F_{lm}^{\rm Out}}$ and 
${D_{lm}^{\rm In}}$ represent
the amplitude of the outgoing waves at infinity and 
the ingoing waves on the horizon 
in the component $\phi_{2}^{\rm SW}$, respectively.
Further, from Eqs.~(\ref{Eq.relDC}) and (\ref{Eq.relDD}) 
it is easy to see the coefficient
${C_{lm}^{\rm In}}$ describing the ingoing waves 
in the component $\phi_{0}^{\rm SW}$ to be
\begin{equation}
 \frac{C_{lm}^{\rm In}}{D^{\rm Out}_{lm}}
  =2\frac{\Gamma(l+1+2iQ)}{\Gamma(l+1-2iQ)}.
  \label{Eq.racd}
\end{equation}
Note that from Eq.~(\ref{Eq.racd}) the ratio 
$|C_{lm}^{\rm In}|^2/|D_{lm}^{\rm Out}|^2$ is given by
\begin{equation}
 |C_{lm}^{\rm In}|^2/|D_{lm}^{\rm Out}|^2=4.
  \label{Eq.relCiDo}
\end{equation}

Next we pay attention to the energy extraction from the black hole.
From Eq.~(\ref{Eq.Erh}), the total flux on the horizon is given by the
component $\phi_{0}$.
On the other hand, the scattered field is induced and connected by the
Kerr-Schild field on the horizon through Eq.~(\ref{eq.detd}).
Further, using Eq.~(\ref{Eq.cond_b}) which requires the absence of the disk
on the horizon, the coefficient $D_{lm}^{\rm Out}$ is obtained as
\begin{equation}
 D_{lm}^{\rm Out}=-ia_{m}\frac{ak_{m}}{r_{1}\Omega_{\rm H}}.
  \label{Eq.Dks}
\end{equation}
Therefore, using the ratio $C_{lm}^{\rm In}/D_{lm}^{\rm Out}$
given by Eq.~(\ref{Eq.racd})
we can evaluate the component $\phi_{0}$ near the horizon as
\begin{equation}
 \phi_{0m}
 \simeq
 \sum_{l\ge|m|}
 e^{-im\omega_{m} t+im\varphi}S^{(1)}_{lm}
 \Delta^{-1}
 e^{-ikr_{*}}
 2\frac{\Gamma(l+1+2iQ)}{\Gamma(l+1-2iQ)}D_{lm}^{\rm Out},
 \label{Eq.phi0sol}
\end{equation}
which is useful to see the distribution of the energy flux 
on the horizon through Eq.~(\ref{Eq.Erh}).

\begin{figure}[h]
 \begin{center}
  \includegraphics*[width=7cm]{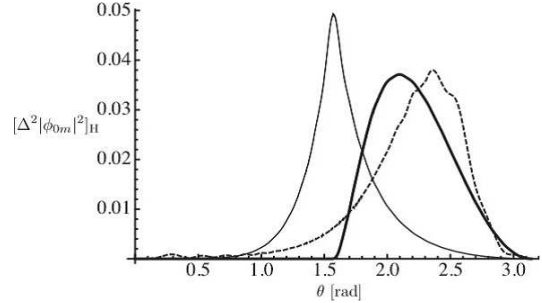}
  \caption{The amplitude $\Delta^2|\phi_{0m}|^2$ of 
  the mode with $m=2$ on
  the horizon as a function of zenithal angle $\theta$. 
  The thin line, dashed line
  and heavy line correspond to the cases $Q=0$ ($a=0$), $Q=10$ ($a\simeq 0.995$),
  and $Q\to \infty$ ($a\to 1$), respectively.
  \label{phihm}}
 \end{center}
\end{figure}

The $\theta$-dependence of $\Delta^2|\phi_{0m}|^2$ 
on the horizon is shown  in Fig.~\ref{phihm},
from which we can see that a highly asymmetric profile appears as the
spin parameter $a$ increases.
For example,
for the modes with $m\ge 2$ the peak of $\Delta^2|\phi_{0m}|^2$ exists in
the lower hemisphere $\pi/2<\theta<\pi$, though the K-S field
$\phi_{2m}^{\rm KS}$ vanishes in this range.
On the contrary, for the modes with $m\le -2$, the peak of
$\Delta^2|\phi_{0m}|^2$ on the horizon exists in the upper hemisphere
$0<\theta<\pi/2$ according to the relation 
$\Delta^2|\phi_{0m}(\theta)|^2=\Delta^2|\phi_{0 -m}(\pi-\theta/2)|^2$.
Even if the net energy flux estimated by the sum 
$\langle{\cal E}^{r}(\theta)\rangle_{\rm H}
\equiv \sum_{m\ge 2} \langle{\cal E}^{r}_{m}(\theta)\rangle_{\rm H}
+ \sum_{m\le -2} \langle{\cal E}^{r}_{m}(\theta)\rangle_{\rm H}$
has a $\theta$-dependence symmetric with respect to the equational
plane, the asymmetric profile of 
$\sum_{m\ge 2} \langle{\cal E}^{r}_{m}(\theta)\rangle_{\rm H}$
given by $\Delta^2|\phi_{0m}(\theta)|^2$ is an interesting feature of
the Kerr black hole.

\begin{figure}[h]
 \begin{center}
  \includegraphics*[width=7cm]{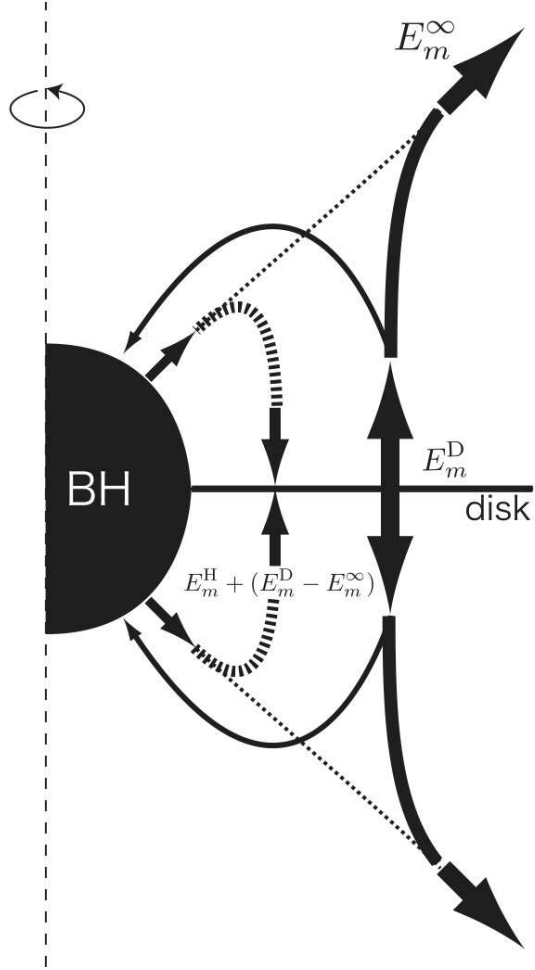}
  \caption{
  Schematic diagram of energy transport in the disk-black hole
  system. The energy radiated from disk surface is shown as solid
  arrows. The superradiant energy flows from the black hole to 
  the disk surface and
  to infinity are shown as heavy dashed and thin dashed arrows, respectively.
  \label{fig2}}
 \end{center}
\end{figure}

Now let us evaluate $E^{{\rm D}}_{m}$ in Eq.~(\ref{Eq.EneDT}) and $E^{\infty}_{m}$ in
Eq.~(\ref{Eq.EneinfT}) under the low-frequency approximation, keeping the terms up to
the first order in $M\omega_{m}$.
The total flux radiated form the disk surface
is given as
\begin{equation}
 E^{\rm D}_{m}
 \simeq
 \frac{(m-a\omega_{m})}{4r_{1}}|a_{m}|^2,
\end{equation}
while the total flux at infinity is obtained by
\begin{equation}
 E^{\infty}_{m}\simeq
  \frac{(m-2a\omega_{m})|a_{m}|^2}{4r_{1}^{\ 2}}.
\end{equation}
The difference
\begin{equation}
 E^{{\rm D}}_{m}-E^{\infty}_{m}
  \simeq
  \frac{a\omega_{m}}{2r_{1}^{\ 2}}
  |a_{m}|^2,
  \label{Eq.diffe}
\end{equation}
means the energy inflow from the disk to the black hole.
On the other hand,
from Eqs.~(\ref{Eq.Tform}), (\ref{Eq.phi0sol}) and (\ref{Eq.relCiDo})
the net extracted energy on the horizon
is calculated as
\begin{equation}
 E^{{\rm H}}_{m}
  \simeq
  \frac{a\omega_{m}}{4r_{1}^{\ 2}}|a_{m}|^2.
  \label{Eq.ehex2}
\end{equation}
Now we can discuss the energy transported in disk-black hole system 
under the low-frequenct approximation.
From Eq.~(\ref{Eq.diffe}) a part of disk emission
turns out to be transported to the black hole.
On the other hand, from Eq.~(\ref{Eq.ehex2}), 
it is easy to see that the net flux on the horizon indicates the energy
extraction from the black hole.
Considering the energy conservation in the disk-black hole system,
the energy flow (\ref{Eq.diffe}) transported from disk to black hole
induces the black hole superradiance (\ref{Eq.ehex2}), and
the total flux $E^{{\rm H}}_{m}+(E^{{\rm D}}_{m}-E^{\infty}_{m})$ returns to
the disk surface (see Fig.~\ref{fig2}). 
Then, it is interesting to note that the feedback energy flow is estimated to be
\begin{equation}
 E^{{\rm H}}_{m}+(E^{{\rm D}}_{m}-E^{\infty}_{m})
  =\frac{3}{2}(E^{{\rm D}}_{m}-E^{\infty}_{m}),
\end{equation}
which is one and half times as large as
the energy flow $E_{m}^{\rm D}-E_{m}^{\infty}$ from disk to black hole.
On the other hand, 
if we consider the limit $a\to 0$ (i.e. the case of the Schwarzschild black hole),
the net flux on the horizon is given by
\begin{equation}
 E^{{\rm H}}_{m}
  \simeq
  -\frac{\omega_{m}^{\ 2}}{4m}|a_{m}|^2,
  \label{Eq.ehex}
\end{equation}
that is,
no energy feedback from the black hole to the disk occurs.

Finally let us discuss the energy flux observed at infinity.
From Eq.~(\ref{Eq.einf}),
the net flux at infinity can be divided as follows,
\begin{equation}
 E_{m}^{\infty}
  =F^{\rm D}_{m}+F^{\rm D-H}_{m}+F^{\rm H}_{m}
  ,
  \label{Eq.diveinf}
\end{equation}
where
\begin{eqnarray}
 F^{\rm D}_{m}&=&
  \int_{0}^{\pi}H_{m}^{\ 2}|a_{m}|^2
  \frac{e^{2a\omega_{m}\cos\theta}\tan^{2m}(\theta/2)}
  {2r_{1}^{\ 2}\sin\theta}
  \left[
   (m-a\omega_{m})^2+\omega_{m}^{\ 2}r_{1}^{\ 2}\cos^2\theta
  \right]{\rm d}\theta,
  \label{Eq.FD-m}
  \\
 F^{\rm D-H}_{m}&=&
  2{\rm Re}
  \left\{
  \int_{0}^{\pi}
   H_{m}\frac{a_{m}e^{a\omega_{m}\cos\theta}\tan^{m}(\theta/2)}
   {\sqrt{2}r_{1}}
   \left[
    (m-a\omega_{m})-i\omega_{m}r_{1}\cos\theta
   \right]
   \sum_{l}
   \bar{F}^{\rm Out}_{lm}
   S^{(-1)}_{lm}(\theta){\rm d}\theta
  \right\},
  \nonumber\\
  \label{Eq.FD-H}
\\
 F^{\rm H}_{m}&=&
  \sum_{ll'}
  \int_{0}^{\pi}
  F_{lm}^{\rm Out}\bar{F}_{l'm}^{\rm Out}
  S^{(-1)}_{lm}(\theta)S^{(-1)}_{l'm}(\theta)
  \sin\theta{\rm d}\theta
  .
\end{eqnarray}
Here from Eqs.~(\ref{Eq.relhinf}) and (\ref{Eq.Dks}) 
the coefficient $F_{lm}^{\rm Out}$ is obtained as
\begin{equation}
 F_{lm}^{\rm Out}
  =
  (-1)^{-l+1}
  a_{m}
  \frac{(l-1)!(l+1)!}{(2l)!(2l+1)!}
  \frac{\Gamma(l+1+2iQ)}{\Gamma(l+2iQ)}
  \frac{2^{l+1}(r_{1}-r_{2})^{l}a\omega_{m}^{\ l+1}k_{m}}{r_{1}\Omega_{\rm H}}
  .
  \label{Eq.CoefF}
\end{equation}
We interpret
$F^{\rm D}_{m}$,
$F^{\rm H}_{m}$ and $F^{\rm D-H}_{m}$ as
the direct radiation from the disk, 
the net flux of the scattered wave caused by
superradiant scattering, 
and their interference effect, respectively.
To find the contribution of the scattered radiation in the net flux at infinity (\ref{Eq.diveinf}),
we pay attention to the dependence of frequency $\omega_{m}$.
From Eqs.~(\ref{Eq.Dks}), (\ref{Eq.diveinf}), (\ref{Eq.FD-H}) and (\ref{Eq.CoefF}), 
it is easy to see that in the interference effect $F^{\rm D-H}_{m}$,
the scattered radiation appear from the order of $(M\omega_{m})^{l+1}$.
However, this term is very smaller than the disk radiation term with the order of
$(M\omega_{m})^0$ in Eq.~(\ref{Eq.FD-m}),
if the low-frequency limit is considered.

From the viewpoint of highly energetic astrophysical phenomena, 
the superradiant transport of the energy flux to the disk will be interesting as a black hole
feedback mechanism which plays a role of disk heating, while the contribution of 
the superradiant scattering of waves to the energy flux at infinity will be useful for an
observational check of the black hole spin (see Fig.~\ref{fig2}).
Unfortunately, as was above-mentioned, the superradiant part at infinity remains 
much smaller than the direct $M\omega_{m}=M m\sigma\ll 1$ is used.
Therefore, it is important to analyze the case such that $M\sigma\sim 1$ 
for the frequency parameter $\sigma$ describing the disk radiation,
by keeping the superradiance condition $\sigma<\Omega_{\rm H}$ and the 
regularity condition $a\sigma<1$ for the Kerr-Schild field in any region except on the disk.

For high $m$ modes giving $m>1 /M \sigma$ the superradiant effect may be also
suppressed. Nevertheless, we must remark that the contribution of such modes
should be taken into account if one estimates the total flux of the disk radiation given
by the function $\psi(X)$ and $\Psi(X)$ singular at the equatorial plane. This is
because as a result of the existence of the branch point at $|X|=1$ the coefficients
$a_{m}$ and $b_{m}$ in the expansion form (\ref{Eq.exp_ps}) and 
(\ref{Eq.exp_P}) do not rapidly decrease as
$m$ increase. Then, the high $m$ modes of the vacuum field given by $\phi_{0}^{\rm SW}$ 
and $\phi_{2}^{\rm SW}$ should be also efficiently generated from the disk radiation 
corresponding to the Kerr-Schild $m$ modes. Even if the superradiant effect is small 
for each high $m$ mode, the total sum (\ref{Eq.expene}) of the energy fluxes 
$\langle {\cal E}^{\mu}_{m}\rangle$ for all $m$ modes will be an important task to discuss
more clearly the energy transport in the disk-black hole system.
Because our main purpose in this paper is focused on the construction of the basic formulae
to calculate of the energy fluxes at the horizon,
the equatorial disk and the far distant region, such a calculation of the total energy flux will
be investigated in future works.

\section*{Acknowledgment}

This research was partially supported by the Grant-in-Aid for Nagoya
University Global COE Program, 
``Quest for Fundamental Principles in the
Universe: from Particles to the Solar System and the Cosmos'', 
from the Ministry
of Education, Culture, Sports, Science and Technology of Japan.

\appendix

\newpage 

\end{document}